\documentclass[12pt]{article}
\usepackage{amsfonts,amsmath,amssymb,epsf,cite}
\usepackage{graphicx,color}
\usepackage{color}
\usepackage{graphicx}% Include figure files
\usepackage{dcolumn}% Align table columns on decimal point
\usepackage{bm}% bold math
\usepackage[hypertex]{hyperref}

\begin{document}

\begin{titlepage}
\thispagestyle{empty}
%\begin{flushright}
%\end{flushright}

\bigskip

\begin{center}
\noindent{\Large Topological Insulators and Superconductors from D-brane}\\
\vspace{2cm} \noindent{
Shinsei Ryu$^{a}$\footnote{e-mail:sryu@berkeley.edu}
and Tadashi Takayanagi$^{b}$\footnote{e-mail:tadashi.takayanagi@ipmu.jp}}

\vspace{1cm}
  {\it
 $^{a}$Department of Physics, University of California, Berkeley, CA 94720, USA

 $^{b}$Institute for the Physics and Mathematics of the Universe (IPMU), \\
 University of Tokyo, Kashiwa, Chiba 277-8582, Japan\\
 }
\end{center}
\vspace{2cm}

\begin{abstract}
Realization of topological insulators (TIs) and superconductors (TSCs),
such as
the quantum spin Hall effect and the $\mathbb{Z}_2$ topological insulator,
in terms of D-branes in string theory is proposed.
We establish a one-to-one correspondence between
the K-theory classification of TIs/TSCs and D-brane charges.
The string theory realization of TIs and TSCs comes naturally with
gauge interactions, and
the Wess-Zumino term of the D-branes gives rise to a gauge field theory
of topological nature.
This sheds light on
TIs and TSCs beyond non-interacting systems,
and the underlying topological field theory
description thereof.
\end{abstract}

\end{titlepage}

\newpage
\section{Introduction}
A gapped state of quantum condensed matter is called
topological phase when it supports stable gapless boundary modes,
such as an edge or a surface state.
The integer quantum Hall effect (QHE),
which exists in $d=2$ spatial dimensions and under a strong magnetic field,
is the best known example of such a phase.
The recent discovery of the quantum spin Hall effect (QSHE) in $d=2$
%\cite{KaneMele,bernevig06,moore07,Roy, konig07}
and the $\mathbb{Z}_2$ topological insulator in $d=3$
\cite{KaneMele,bernevig06,moore07,Roy,konig07, Fu06_3Da, hasan07, Qi2008}
shows topological phases can exist even in $d>2$ spatial dimensions,
and can be protected by some discrete symmetries
such as time-reversal symmetry (TRS, T),
particle-hole symmetry (PHS, C),
and
chiral (or sublattice) symmetry (SLS, S).

For non-interacting fermions,
an exhaustive classification of topological insulators (TIs) and superconductors (TSCs)
is proposed in Refs.\ \cite{SRFL,Kitaev}:
TIs/TSCs are classified in terms of spatial dimensions $d$
and the $10=2+8$ symmetry classes
(two ``complex'' and eight ``real'' classes)
(Table \ref{charge}).
The ten symmetry classes are in one-to-one correspondence
to the Riemannian symmetric spaces (without exceptional series) and,
as pointed out in \cite{Kitaev}, they are equivalent to
K-theory classifying spaces \cite{Horava}.
For example, the IQHE, QSHE,
and $\mathbb{Z}_2$ TI
are a topologically non-trivial state
belonging to class A ($d=2$), AII ($d=2$), and AII ($d=3$), respectively.

The complete classification of non-interacting TIs and TSCs opens up a number of further questions,
most interesting among which are interaction effects:
%To list a few key questions:
Do non-interacting topological phases continue to exist
in the presence of interactions?
Can interactions give rise to novel topological phases other than non-interacting TIs/TSCs?
%Is there (interacting) a TI/TSC,
%which support fractionalized excitations,
%other than the fractional QHE?
%(e.g., is there ``the fractional QSHE''?)
What is a topological field theory underlying TIs/TSCs, which can potentially
describe TIs/TSCs beyond non-interacting examples?, etc.

On the other hand, the ten-fold classification of TIs/TSCs reminds us of D-branes,
which are fundamental objects in string theory,
and are also classified by K-theory \cite{WittenK}
(Table \ref{dbrane}) via the open string tachyon condensation \cite{Sen}.
It is then natural to speculate a possible connection between TIs/TSCs and of D-branes.
In this paper, we propose a systematic construction of TIs/TSCs
in terms of two D-branes (D$p$- and D$q$-branes),
possibly with an orientifold plane (O-plane).
Besides the appealing mathematical similarity between TIs/TSCs and D-branes,
realizing TIs/TSCs in string theory has a number of merits, since
string theory and D-branes are believed
to be rich enough to reproduce many types of field theories and interactions
in a fully consistent and UV complete way.
Indeed, our string theory realizations of TIs/TSCs give rise to
massive fermion spectra,
which are in one-to-one correspondence with
the ten-fold classification of TIs/TSCs,
and come quite naturally
with gauge interactions.
These systems, while interacting,
are all topologically stable, as protected by the K-theory charge of D-branes.
We thus make a first step toward understanding interacting TIs/TSCs
\cite{Karch}. We are also separately preparing a regular paper with more details and
expanded results in \cite{RTTA}.

In D$p$-D$q$-systems,
massive fermions arise as an open string excitation between the two D-branes.
The distance between the branes corresponds to the mass of fermions.
Open strings ending on the same D-branes give rise to a gauge field,
which we call $A_\mu$ (D$p$) and $\tilde{A}_\mu$ (D$q$)
with gauge group $G$ and $\tilde{G}$, respectively,
and couple to the fermions.
These two gauge fields play different roles in our construction:
The gauge field $A_\mu$ ``measures'' K-theory charge of the D$q$-brane,
and in that sense it can be interpreted as an ``external'' gauge field.
In this picture, the D$q$-brane charge is identified with the
topological (K-theory) charge of TIs/TSCs.
On the other hand,
$\tilde{A}_\mu$ is an internal degree of freedom on the D$q$-brane.
%and can be viewed as an ``internal'' gauge field.
For example, in the integer/fractional QHE,
the external gauge field is the electromagnetic U(1) gauge field,
which measures the Hall conductivity,
while the internal gauge field is the Chern-Simons (CS) gauge field describing
the dynamics of the droplet itself.

The massive fermions can be integrated out, yielding
the description of the topological phase in terms of the gauge fields.
The resulting effective field theory comes with terms of topological nature,
such as the CS or the $\theta$-terms.
In our string theory setup,
they can be read off from the Wess-Zumino (WZ) action of the D-branes,
by taking one of the D-branes as a background for the other.
One can view
these gauge-interacting TIs/TSCs from D$p$-D$q$-systems
as an analogue of
the projective (parton) construction of the (fractional) QHE
\cite{Wen91-99}.
%where a topological field theory description for a given
%trial wave function of the FQHE is constructed
%by considering fermions (or bosons) in an integer quantum
%Hall state interacting through internal gauge fields.
%By integrating gapped fermions (or bosons) yields
%a topological field theory of the Chern-Simons type.
Our string theory realization of TIs/TSCs
sheds light on extending the projective construction of
the QHE to more generic TIs/TSCs;
it tells us what type of gauge field is ``natural'' to couple with fermions
in topological phases, and guarantees the topological stability
of the system.
%In a sense, aided by string theory,
%we extend in this paper the parton construction of the factional QHE
%to all TIs/TSCs in the ten-fold classification.
%(Our construction does not realize, however, the fractionalized
%TIs/TSCs, at least naively.)
%\textcolor{red}{
%While the description of response of TIs to external gauge field in terms
%of topological field theories is developed in Ref.\ \cite{Qi2008},
%the description of the dynamics of TIs/TSCs
%in terms of (topological) field theory is still lacking.}

\begin{table}
\begin{center}
\begin{tabular}{|c|cccccccc|ccc|}\hline
   $\mathrm{class} \backslash d$  & 0 & 1 & 2 & 3 & 4 & 5 & 6 & 7 & T & C & S  \\  \hline
  A & $\mathbb{Z}$ & 0 & $\mathbb{Z}$ & 0 & $\mathbb{Z}$ & 0 & $\mathbb{Z}$ & 0             & 0 & 0 & 0    \\
  AIII & 0 & $\mathbb{Z}$ & 0 & $\mathbb{Z}$ & 0 & $\mathbb{Z}$ & 0 & $\mathbb{Z}$          & 0 & 0 & 1    \\  \hline
  AI & $\mathbb{Z}$ & 0 & 0 & 0 & $2\mathbb{Z}$ & 0 & $\mathbb{Z}_2$ & $\mathbb{Z}_2$    & $+$ & 0 & 0     \\
  BDI & $\mathbb{Z}_2$ & $\mathbb{Z}$ & 0 & 0 & 0 & $2\mathbb{Z}$ & 0 & $\mathbb{Z}_2$     & $+$ & $+$ & 1    \\
  D & $\mathbb{Z}_2$ & $\mathbb{Z}_2$ & $\mathbb{Z}$ & 0 & 0 & 0 & $2\mathbb{Z}$ & 0     & 0 & $+$ & 0     \\
  DIII & 0 & $\mathbb{Z}_2$ & $\mathbb{Z}_2$ & $\mathbb{Z}$ & 0 & 0 & 0 & $2\mathbb{Z}$  & $-$ & $+$ & 1     \\
  AII & $2\mathbb{Z}$ & 0 & $\mathbb{Z}_2$ & $\mathbb{Z}_2$ & $\mathbb{Z}$ & 0 & 0 & 0   & $-$ & 0 & 0     \\
  CII & 0 & $2\mathbb{Z}$ & 0 & $\mathbb{Z}_2$ & $\mathbb{Z}_2$ & $\mathbb{Z}$ & 0 & 0   & $-$ & $-$ & 1     \\
  C & 0 & 0 & $2\mathbb{Z}$ & 0 & $\mathbb{Z}_2$ & $\mathbb{Z}_2$ & $\mathbb{Z}$ & 0     & 0 & $-$ & 0     \\
  CI & 0 & 0 & 0 & $2\mathbb{Z}$ & 0 & $\mathbb{Z}_2$ & $\mathbb{Z}_2$ & $\mathbb{Z}$    & $+$ & $-$ & 1
   \\ \hline
\end{tabular}
\end{center}
\caption{
\label{charge}
Classification of topological insulators and superconductors
\cite{SRFL,Kitaev};
$d$ is the space dimension;
the left-most column (A, AIII, $\ldots$, CI)
denotes the ten symmetry classes of fermionic Hamiltonians,
which are characterized by the presence/absence of
time-reversal (T), particle-hole (C), and chiral (or sublattice) (S) symmetries
of different types denoted by $\pm 1$ in the right most three columns.
The entries ``$\mathbb{Z}$'', ``$\mathbb{Z}_2$'', ``$2\mathbb{Z}$'',
and ``$0$'' represent the presence/absence of
topological insulators and superconductors,
and when they exist, types of these states
(see Ref.\ \cite{SRFL} for detailed descriptions).
}
\end{table}

\begin{table}
\begin{center}
\begin{tabular}{|cccccccccccc|}\hline
    & D($-1$) & D0 & D1 & D2 & D3 & D4 & D5 & D6 & D7 & D8 & D9 \\ \hline
  type IIB & $\mathbb{Z}$ & 0 & $\mathbb{Z}$ & 0 & $\mathbb{Z}$ & 0 & $\mathbb{Z}$ & 0 & $\mathbb{Z}$ & 0 & $\mathbb{Z}$ \\
  O9$^{-}$ (type I) & $\mathbb{Z}_2$ & $\mathbb{Z}_2$ & $\mathbb{Z}$ & 0 & 0 & 0 & $\mathbb{Z}$ & 0 & $\mathbb{Z}_2$ & $\mathbb{Z}_2$ & $\mathbb{Z}$ \\
  O9$^{+}$ & 0 & 0 & $\mathbb{Z}$ & 0 & $\mathbb{Z}_2$ & $\mathbb{Z}_2$ & $\mathbb{Z}$ & 0 & 0 & 0 & $\mathbb{Z}$ \\ \hline
\end{tabular}
\end{center}
\caption{\label{dbrane}
D$p$-brane charges from K-theory, classified by
$\mathrm{K}(\mathbb{S}^{9-p})$,
$\mathrm{KO}(\mathbb{S}^{9-p})$ and
$\mathrm{KSp}(\mathbb{S}^{9-p})$ \cite{WittenK}.
A $\mathbb{Z}_2$ charged D$p$-brane with
$p$ even or $p$ odd represents
a non-BPS D$p$-brane
or a bound state of a D$p$ and an anti-D$p$ brane, respectively \cite{Sen}.
}
\end{table}

\begin{table}
\begin{center}
\begin{tabular}{|c|c|cccccccc|}\hline
 $G$  &  $\mathrm{class}\backslash d$  & 0 & 1 & 2 & 3 & 4 & 5 & 6 & 7    \\  \hline
U & A   & U & - & U & - & U & - & U & -    \\
U & AIII& - & U & - & U & - & U & - & U     \\  \hline
O & AI  & O & - & - & - &Sp & - & U & O       \\
O & BDI & O & O & - & - & - & Sp& - & U      \\
O & D   & U & O & O & - & - & - & Sp& -       \\
O & DIII& - & U & O & O & - & - & - & Sp      \\
Sp& AII &Sp & - & U & O & O & - & - & -       \\
Sp& CII & - &Sp & - & U & O & O & - & -       \\
Sp& C   & - & - &Sp & - & U & O & O & -       \\
Sp& CI  & - & - & - &Sp & - & U & O & O
   \\ \hline
\end{tabular}
\end{center}
\caption{\label{gauge}
External $G$ (left-most column) and internal $\tilde{G}$ gauge groups
for each spatial dimension $d$ and symmetry class;
U, O, Sp, represents U(1), $\mathrm{O}(1)=\mathbb{Z}_2$,
and $\mathrm{Sp}(1)=\mathrm{SU}(2)$,
respectively.
}
\end{table}

%\paragraph{Complex case}

\section{Complex case}
Let us start with the most familiar example of the QHE (class A in $d=2$).
We fix the value of $p$ to be $p=5$ by T-duality, and
consider a D5-brane in type IIB string theory which extends in the
$x^{0,1,2,3,4,5}$ directions in ten-dimensional space-time.
We take the D$q$-brane with $q=5$ in the $x^{0,1,2,6,7,8}$ directions (Table \ref{AandAIII}).
By T-duality, this setup is equivalent to
the D3-D7 system studied in \cite{Rey,DKS,FujitaFQHE}.
Since the number of Neumann-Dirichlet (ND) directions is six,
open string excitations between the D5-branes give rise to
two Majorana fermions (Mj)
[= one two-component Dirac fermion (Di), $\psi$] and no bosons.
The distance between the D-branes in $x^9$ direction ($\Delta x^9$)
is proportional
to the mass $m$ of the fermions.
The low-energy effective theory is schematically summarized by the
effective Lagrangian in the $(2+1)$-dimensional common direction of the two D5-branes,
\begin{eqnarray}
%$
\mathcal{L}
=
\bar{\psi}
[
\gamma^{\mu}
(
i\partial_{\mu} - A_{\mu} -\tilde{A}_{\mu}
)
- m
]
\psi
+
\cdots.
%$
\end{eqnarray}
Integrating the massive fermions yields the CS terms
$\frac{k}{4\pi}\int A\wedge dA$ and $\frac{k}{4\pi}\int \tilde{A}\wedge d\tilde{A}$
with $k=\pm 1/2$ (parity anomaly).
The Hall conductivity is read off from the CS term for $A_{\mu}$
as $\sigma_{xy}=k/(2\pi)$.
Alternatively,
the presence of the CS terms can be read off from
the WZ action of either one of D5-branes,
e.g.,
%$
\begin{eqnarray}
S_{\mathrm{D}5}^{\mathrm{WZ}}\propto\int_{\mathrm{D}5}F\wedge F\wedge C_{2}
=\int_{\mathrm{D}5}A\wedge F\wedge (dC)_{3}
\end{eqnarray}
%$
for the external gauge field $A_{\mu}$,
where $C_2$ is the RR 2-form from the D$q$-brane.
%Similarly,
%the CS action for $\tilde{A}_{\mu}$ is derived
%from the DBI action for the D5-brane
%via the Wess-Zumino coupling.
When we change the sign of $m$
by passing the D$q$-brane through the D$p$-brane,
the value of $k$ jumps from $\pm 1/2$ to $\mp 1/2$.
If we instead put $N_f$ D$q$-branes,
we have $N_f$ copies of massive Dirac fermions $\psi_i$
which couple with $\mathrm{U}(N_f)$ gauge fields
$A_{\mu}$ and $\tilde{A}_{\mu}$
(when all D$q$ are coincident).

This brane construction
can be extended to other even space dimensions $d=2n$
by considering D$5$-D$q$ systems with $q=5,7$ (Table \ref{AandAIII}).
This setup gives rise to
the fermion spectrum consisting of one Dirac fermion per D$q$-brane,
and the CS terms of the form $\propto k\int A\wedge F^{n}$.
All of these brane configurations are identified with
class A TIs, which are characterized by the absence of any discrete symmetries.

Now let us turn to AIII,
which is characterized by the presence of SLS.
We argue that SLS is equivalent
to an invariance of the brane configurations
under inversion of a coordinate in the Dirichlet-Dirichlet (DD) directions
of open strings between the D$5$- and D$q$-branes.
One way to realize this is to assume two DD directions,
say, $x^1$ and $x^9$, and impose $x^1=0$.
Indeed such a configuration is obtained by taking T-dual
of class A configurations (Table \ref{AandAIII}).
Again,
the fermion spectrum consists of two Majorana fermions (=one Dirac fermion)
for all dimensions, and
the mass of the fermion is, again, proportional to $\Delta x^9$.

In our setup in general,
the number of Neumann-Neumann (NN) directions $\#\mathrm{NN}$
is equal to the space-time dimensions $d+1$ of TIs/TSCs.
On the other hand, $\#\mathrm{DD}$ represents the number
of possible mass deformations:
it is one if there is no SLS (class A),
while it is two in the presence of SLS.
Finally, $\#\mathrm{ND}$ is determined by $\#\mathrm{NN}$ and
$\#\mathrm{DD}$ via the relation $\#\mathrm{ND}=10-\#\mathrm{NN}-\# \mathrm{DD}$.
Note also that the T-dual in any ND directions does not change
$\#\mathrm{NN}$, $\#\mathrm{DD}$ and $\#\mathrm{ND}$ and thus
is a redundant operation.

\begin{table}
\begin{center}
\begin{tabular}{|c|cccccc|cccc|c|c|}\hline
   & $0$ & $1$ & $2$ & $3$ & $4$ & $5$ & $6$ & $7$ & $8$ & $9$ &  $d$ & A \\ \hline
 D5 & $\times$ &  $\times$ & $\times$ & $\times$ & $\times$ &  $\times$ &   &   &   &   &  &   \\ \hline
 D3 & $\times$ &  &  &  &  &        & $\times$ & $\times$ & $\times$ &  &   0 & $\mathbb{Z}$ (2 Mj)  \\
 D5 & $\times$ & $\times$ & $\times$ &  &  &       & $\times$ & $\times$ & $\times$ &   & 2 & $\mathbb{Z}$ (2 Mj)  \\
 D7 & $\times$ & $\times$ & $\times$ & $\times$ & $\times$  &     & $\times$ & $\times$ & $\times$ &   &4 & $\mathbb{Z}$ (1 Di) \\  \hline
  \end{tabular}
\end{center}
\begin{center}
\begin{tabular}{|c|cccccc|ccccc|c|c|}\hline
       & $0$ & $1$ & $2$ & $3$ & $4$ & $5$ & $6$ & $7$ & $8$ & $9$ & & $d$ & AIII  \\ \hline
 D4 & $\times$ &   & $\times$ & $\times$ & $\times$ &  $\times$ &   &   &   &   &  & & \\  \hline
 D4 & $\times$ &  & $\times$ &  &  &        & $\times$ & $\times$ & $\times$ &  & & 1 & $\mathbb{Z}$ (2 Mj) \\
 D6 & $\times$ &   & $\times$ & $\times$ & $\times$ &       & $\times$ & $\times$ & $\times$ &  & &3&   $\mathbb{Z}$ (2 Mj) \\ \hline
  \end{tabular}
\end{center}
\caption{
\label{AandAIII}
D$p$-D$q$ systems for class A and AIII
where $p=5$ and $q=3,5,7$ for A,
and $p=4$ and $q=4,6$ for AIII.
The D-branes extend in the $\mu$-th direction denoted by ``$\times$''
in the ten-dimensional space-time ($\mu=0,\ldots,9$);
$d+1$ is the number of common directions of D$p$-and D$q$-branes;
The last column shows the D$q$-brane charge,
together with fermion spectra per D$q$-brane,
where  "$N_f$ Mj'' or ``$N_f$ Di''
represents $N_f$ flavor of Majorana and Dirac spinor, respectively.}
\end{table}

%\paragraph{Real case}

\section{Real case}
To realize eight ``real'' symmetry classes in string theory,
we need to implement TRS and PHS.
To preserve PHS, we require the internal gauge field $\tilde{A}_\mu$
is not independent of its complex conjugate.
This is the same as the orientation ($\Omega$) projection in string theory.
%Note that there are two
%different $\Omega$ projections: orthogonal (O) and symplectic (Sp) types.
To realize TRS, we recall that it can be viewed as a product of PHS and SLS
\cite{SRFL}.
As SLS can be imposed as a parity symmetry in string theory,
we can interpret TRS as the orientifold projection.

Let us start with class C and D, which are characterized by
the presence of PHS but lack of TRS.
We take an $\Omega$ projection of the class A setup.
Note that there are two types of $\Omega$ projections,
represented by two types of O9-plane, i.e.,
O9$^-$ (orthogonal) and O9$^+$ (symplectic).
While only O$9^-$ leads to supersymmetric type I string theory,
here we consider both because the T-dual of O9$^+$
is equivalent to O$p^+$ planes ($p\leq 8$) in type II string theory.
We realize class C and class D TSCs by
considering a D5-brane which extends in the $x^{0,1,2,3,4,5}$ directions,
in the presence of an O9-plane.
As before, we put a D$q$-brane with $q=d+3$ so that there are
$d+1$ common directions (Table \ref{real case}).
For class AII and AI,
characterized by the presence of TRS but lack of PHS,
we take the orientifold projection which leads to an O8-plane in type IIA theory \cite{ISS}.
By choosing
$(p,q)=(4,d+4)$,
%$p=4$ and $q=d+4$,
we obtain the brane configuration given
in the third table in Table \ref{real case}.
Though the D-brane charges with
an O$p$-plane for $p\leq 8$ are classified by KR-theory, the same result
can be obtained from KO-theory via T-duality for our purpose \cite{ISS}.
Finally, the remaining four classes,
CII, BDI, CI and DIII
can be obtained by taking DD directions
to be two instead of one.
SLS is imposed by requiring $x^9=0$ for all of these classes.
These are O8- or O9-projection of the class AIII setup
(Table \ref{real case}).

\begin{table}
\begin{center}
\begin{tabular}{|c|cccccc|cccc|c|c|c|}\hline
   & $0$ & $1$ & $2$ & $3$ & $4$ & $5$ & $6$ & $7$ & $8$ & $9$& $d$  & C (O$9^{-}$) & D (O$9^{+}$) \\ \hline
 D5 & $\times$ & $\times$  & $\times$ & $\times$ & $\times$ & $\times$       &  &  &  &  &  &  & \\ \hline
 D3 & $\times$ &   &  &  &  &        & $\times$ & $\times$ & $\times$ &   & 0 & 0 & $\mathbb{Z}_2$ (2 Mj)\\
 D4 & $\times$ & $\times$ &   &  &  &       & $\times$ & $\times$ & $\times$ & & 1 & 0 &  $\mathbb{Z}_2$ (1 Mj) \\
 D5 & $\times$ & $\times$ & $\times$ &   &   &     & $\times$ & $\times$ & $\times$ &  & 2 &  $\mathbb{Z}$ (4 Mj)  &  $\mathbb{Z}$ (1 Mj) \\
 D6 & $\times$ & $\times$  & $\times$ & $\times$ &  &        & $\times$ & $\times$ & $\times$ & & 3  & 0 & 0 \\
 D7 & $\times$ & $\times$ & $\times$ & $\times$ & $\times$ &        & $\times$ & $\times$ & $\times$ & & 4  & $\mathbb{Z}_2$ (2 Di) & 0 \\ \hline
  \end{tabular}
\end{center}
\begin{center}
\begin{tabular}{|c|cccccc|cccc|c|c|c|}\hline
   & $0$ & $1$ & $2$ & $3$ & $4$ & $5$ & $6$ & $7$ & $8$ & $9$ & $d$ & CI (O$9^{-}$) & DIII (O$9^{+}$) \\ \hline
   D5 & $\times$ & $\times$  & $\times$ & $\times$ & $\times$ & $\times$       &  &  & & &  & &  \\ \hline
 D2 & $\times$ &   &  &  &  &        & $\times$ & $\times$ &  & &0  & 0 & 0 \\
 D3 & $\times$ & $\times$ &   &  &  &       & $\times$ & $\times$ &  & &1  & 0 &  $\mathbb{Z}_2$ (2 Mj) \\
 D4 & $\times$ & $\times$ & $\times$ &   &   &     & $\times$ & $\times$ &  & &2  & 0 & $\mathbb{Z}_2$ (2 Mj) \\
 D5 & $\times$ & $\times$  & $\times$ & $\times$ &  &        & $\times$ & $\times$ &  & &3   &  $\mathbb{Z}$ (4 Mj)  & $\mathbb{Z}$ (1 Mj) \\ \hline
  \end{tabular}
\end{center}
\begin{center}
\begin{tabular}{|c|ccccc|c|cccc|c|c|c|}\hline
    &$0$& $1$ & $2$ & $3$ & $4$ & $5$ & $6$ & $7$ & $8$ & $9$ & $d$ & AII (O$8^{-}$)  & AI (O$8^{+}$) \\ \hline
D4&$\times$&$\times$&$\times$&$\times$ &$\times$&        &  &  &  &  &  &  &  \\ \hline
D4&$\times$&            &            &              &             &        &$\times$&$\times$&$\times$&$\times$& 0 & $\mathbb{Z}$ (4 Mj) & $\mathbb{Z}$ (1 Mj)\\
D5&$\times$&$\times$&            &  &  &       & $\times$ & $\times$ & $\times$ & $\times$ & 1 &  0 &  0 \\
D6&$\times$&$\times$&$\times$&   &   &     & $\times$ & $\times$ & $\times$ & $\times$ & 2 &  $\mathbb{Z}_2$ (4 Mj) & 0 \\
D7&$\times$&$\times$&$\times$& $\times$ &   &     & $\times$ & $\times$ & $\times$ & $\times$ & 3 &  $\mathbb{Z}_2$ (2 Mj) & 0 \\
D8&$\times$&$\times$&$\times$& $\times$ & $\times$ &        & $\times$ & $\times$ & $\times$ &  $\times$ & 4 & $\mathbb{Z}$ (1 Di) & $\mathbb{Z}$ (1 Di) \\ \hline
  \end{tabular}
\end{center}
\begin{center}
\begin{tabular}{|c|ccccc|c|cccc|c|c|c|}\hline
 & $0$ & $1$ & $2$ & $3$ & $4$ & $5$ & $6$ & $7$ & $8$ & $9$ & $d$ & CII (O$8^{-}$)  & BDI (O$8^{+}$) \\ \hline
D4 & $\times$ & $\times$  & $\times$ & $\times$ & $\times$ &        &  &  &  &  & &  & \\ \hline
D3 & $\times$ &   &  &  &  &        & $\times$ & $\times$ & $\times$ &   & 0 & 0 & $\mathbb{Z}_2$ (2 Mj)\\
D4 & $\times$ & $\times$ &   &  &  &       & $\times$ & $\times$ & $\times$ &  & 1 &  $\mathbb{Z}$ (4 Mj) &  $\mathbb{Z}$ (1 Mj) \\
D5 & $\times$ & $\times$ & $\times$ &   &   &     & $\times$ & $\times$ & $\times$ & & 2 &  0 & 0 \\
D6 & $\times$ & $\times$  & $\times$ & $\times$ &  &        & $\times$ & $\times$ & $\times$ &  & 3 & $\mathbb{Z}_2$ (4 Mj) & 0 \\
D7 & $\times$ & $\times$  & $\times$ & $\times$ & $\times$  &        & $\times$ & $\times$ & $\times$ &  & 4 & $\mathbb{Z}_2$ (2 Di) & 0 \\ \hline
  \end{tabular}
\end{center}
\caption{
D$p$-D$q$ systems for eight ``real'' symmetry classes,
where $p=5$ for classes C, D, CI, DIII,
and $p=4$ for classes AII, AI, CII, BDI.
For classes AII, AI, CII, BDI,
the O8-plane extends except $x^5$.
\label{real case}
}
\end{table}

For these D-brane configurations,
we chose a D$p$-brane to be
a standard one with the integer K-theory charge so that it is
regarded as the background (bulk) material itself. Then, we find that the K-theory charge of the D$q$-branes
(shown in the last two columns in Table \ref{real case})
agrees precisely with the corresponding classification of TIs/TSCs
(Table \ref{charge}).
Moreover,
the fermion content of these string theory realizations
(denoted in the last two columns in Table \ref{real case}
either by ``$N_f$ Mj'' or ``$N_f$ Di'' with $N_f$ an integer)
can be compared with
the Dirac representative of TIs/TSCs constructed in Ref.\ \cite{SRFL}.
Indeed, they agree completely.
It is also interesting to note that,
for the Dirac representative of TIs/TSCs,
the \textit{momentum} dependence of
the projection operator,
which is one of key ingredients in the classification of TIs/TSCs\cite{SRFL},
looks quite similar to
the spatial profile of the tachyon field
in string theory
in \textit{real} space \cite{WittenK}.

We now describe the field theory content of the
D$p$-D$q$ systems charges in more details.
First, for the $\mathbb{Z}$ TIs/TSCs on the diagonal
in Table \ref{charge} (``primary series''),
the internal gauge group is $\mathrm{O}(1)=\mathbb{Z}_2$.
In particular, for class D in $d=2$,
our string theory realization corresponds to
(a proper supersymmetric generalization of)
the honeycomb lattice Kitaev model in the weak pairing (non-Abelian) phase
\cite{Kitaev05}.
Similarly, for class DIII in $d=3$,
it corresponds to an interacting bosonic model on the diamond lattice
\cite{Ryu08}.

For $\mathbb{Z}_2$ TIs/TSCs of the first descendant
of the primary series, i.e,
BDI ($d=0$), D ($d=1$), DIII ($d=2$), and AII ($d=3$),
the internal gauge group is $\mathrm{O}(1)=\mathbb{Z}_2$
(Table \ref{gauge}).

For $\mathbb{Z}_2$ TIs/TSCs of the second descendant
of the the primary series, i.e,
D ($d=0$), DIII ($d=1$), AII ($d=2$), and CII ($d=3$),
the internal gauge group is U(1).
For D and DIII,
the U(1) unitary gauge field
couples to two real fermions
which can be combined into a single complex field.
For AII and CII, the fermion spectrum consists of 4 Mj.
In this case,
the external gauge field is $\mathrm{Sp}(1)=\mathrm{SU}(2)$,
which couples to a doublet, $\psi_{\uparrow/\downarrow}$.
Each has 2 Mj (=1Di) degrees of freedom and
couples to a U(1) internal gauge field as follows:
\begin{eqnarray}
%$
\mathcal{L}
=
\bar{\psi}
[
\gamma^{\mu}(
i\partial_{\mu} - A_{\mu} -\tilde{A}_{\mu}
)
- m M
] \psi
+\cdots,
%$
\end{eqnarray}
where $\psi= (\psi_{\uparrow},\psi_{\downarrow})^T$,
and
$M$ is a diagonal mass matrix
whose eigenvalue is $\pm 1$ for $\psi_{\uparrow/\downarrow}$,
respectively.

Finally,
for TIs/TSCs labeled by $2\mathbb{Z}$, i.e.,
AII ($d=0$), CII ($d=1$), C ($d=2$), and CI ($d=3$),
the gauge group is $G \times \tilde{G} = \mathrm{SU}(2)\times \mathrm{SU}(2)$,
with 4 Mj fermions in bi-fundamental.

Even though the ten-dimensional string theories in the bulk are supersymmetric,
our brane setups are not in general.
When $\#\mathrm{ND}=4$ with $\mathbb{Z}$ charge,
they exceptionally preserve a quarter of supersymmetries.
In general, when $\#\mathrm{ND}=4$, there exist massive bosons
in addition to the massive fermions. This happens
when $d=4$ for A, C, AI, AII, and when $d=3$ for AIII, CII, CI and DIII.
Since $\#\mathrm{NN}>4$ for all the other branes systems,
we only have fermions from open strings between the D$p$- and D$q$-branes
and there are no tachyons.

We can take the T-duality further in NN directions.
However, this lead to theories with different properties than TIs/TSCs
(Table \ref{gen}).
Note that we have succeeded to realize all TIs/TSCs
in space dimensions $d\leq 4$.

%\paragraph{Boundary of TIs/TSCs}
\section{Boundary of TIs/TSCs}
\begin{figure}
 \begin{center}
 \includegraphics[width=7cm,clip]{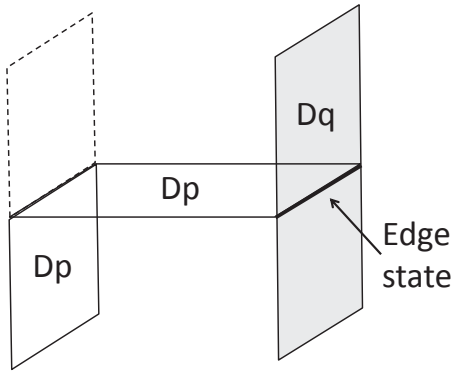}
 \end{center}
 \caption{
 \label{edgestate}
A boundary of a TI/TSC as a brane intersection.
 }
 \end{figure}

A defining property of TIs/TSCs is the appearance of
stable gapless degrees of freedom,
when the system is terminated by a $(d-1)$-dimensional boundary.
In our brane construction,
the sample boundary can be constructed
by bending the D$p$-brane toward the D$q$-brane,
to create an intersection between these branes
(Fig.\ \ref{edgestate}).
%along a $d-1$ dimensional boundary,
This leads to a position-dependent fermion mass,
which changes its sign at the intersection.
%\textcolor{red}{For example, in class A,
%this amounts by replacing $x^1$ with $x^9$ of the
%D$5$(=D$p$)-brane. }
This increases $\#\mathrm{ND}$ by two and the correct number of massless fermions
appears at the intersection.

% In realistic systems, we need to versify that the gapless fermion is stable against the
% random perturbations with the required symmetries. Indeed, the classification into the ten classes has been
% obtained for the random Hamiltonian \cite{BL}. One way to deal with the randomness is to introduce a supersymmetry, which
% leads to bosonic superpartners (or ghosts) for the fermions in TIs/TSCs.
% In our brane setup, this can be done by replacing a D$q$-brane with $m$ D$q$ and $m$ ghost D$q$-branes introduced in
% \cite{OT}, which leads to $U(m|m)$ or $OSp(m|m)$ symmetries.

\begin{table}[t]
\begin{center}
\begin{tabular}{|c|c|c|c|c|}\hline
  $\#\mathrm{DD}$ & (O$9^{-}$,O$9^+$) & (O$8^{-}$,O$8^+$) & (O$7^{-}$,O$7^+$)
  & (O$6^{-}$,O$6^+$)  \\ \hline
  0 & Chiral &  &   &    \\
  1 & (C,D) &  (AII,AI) &  &     \\
  2 & (CI,DIII) & (CII,BDI) & (DIII,CI) &    \\
  3 & $d\leq 2$ & $d\leq 2$ & $d\leq 2$ & $d\leq 2$     \\
  4 & $d\leq 1$ & $d\leq 1$ & $d\leq 1$ & $d\leq 1$     \\  \hline
  \end{tabular}
\end{center}
\caption{
\label{gen}
D$p$-D$q$ systems in the presence of an O-plane;
``$d\ge 2$'' means
the constraint of possible spatial dimensions $d$
in the brane systems due to the existence of
open string tachyons. ``Chiral'' denotes the existence of chiral fermions
and
is interpreted as boundary (edge) states.
The horizontal direction is shifted by a T-duality in the NN direction.
}
\end{table}

\section{Conclusions}

The main conclusion of this paper is that
there is a one-to-one correspondence between the
tachyon-free D$p$-D$q$ systems and the ten classes of topological insulators
in $d\leq 4$ dimensions. Indeed, we explicitly constructed
the corresponding ten classes of D$p$-D$q$ brane configurations
in superstring theory. Two out of ten
are realized in type II string theory without orientifolds, while
the other eight require orientifolds. The K-theory charges of the
D$q$-branes agree with that of the topological insulators.

One may wonder if there are
other tachyon-free D$p$-D$q$ systems which have not been considered in this paper.
However, it turns out that their low-energy theories just correspond to multiple copies of
the ten classes of topological insulators (for details refer to \cite{RTTA}).

Since a topological insulator has a mass gap in the bulk, the distance
between the D$p$ and D$q$ is taken to be non-vanishing in its corresponding D-brane system.
%However, once it has a boundary (or edge state),
When we discuss the boundary of a TI/TSC, however,
the D$p$-brane is bent toward the D$q$-brane and thus D$p$ and D$q$ are intersecting with each other.
Therefore massless fermions
appear at the intersection and they are identified with the boundary modes (or edge modes).

We can also consider holographic descriptions of these systems by extending
the constructions in \cite{DKS,FujitaFQHE} in principle,
though it is not possible to take the large-$N$ limit of
$\mathbb{Z}_2$ charged D-branes.

 \paragraph{Acknowledgments}
We acknowledge
 ``Quantum Criticality and the AdS/CFT correspondence''
 miniprogram at KITP,
 ``Quantum Theory and Symmetries''
 conference at University of Kentucky.
We would like to thank
K.\ Hori, P.\ Kraus, A.\ Ludwig, J.\ Moore,
M.\ Oshikawa, and S.\ Sugimoto
for useful discussion,
and A. Furusaki for his clear lecture at
``Development of Quantum Field Theory and String Theory'' (YITP-W-09-04)
at Kyoto University.
% TT is very grateful to YITP
% at Kyoto University where the workshop (YITP-W-09-04)
% on ``Development of Quantum Field Theory and String Theory'' is held and to
% A. Furusaki for his clear lecture there.
 SR thanks Center for Condensed Matter Theory at University of
 California, Berkeley for its support.
 TT is supported in part by JSPS Grant-in-Aid
 for Scientific Research No.\ 20740132, and
 by JSPS Grant-in-Aid for Creative Scientific Research No.\ 19GS0219.


\begin{thebibliography}{99}

\bibitem{KaneMele}
C.\ L.\ Kane and E.\ J.\ Mele,
Phys.\ Rev.\ Lett.\ \textbf{95}, 146802 (2005);
\textbf{95}, 226801 (2005).

\bibitem{Roy}
R.\ Roy,
Phys.\ Rev.\ B \textbf{79}, 195321 (2009);
\textit{ibid}, 195322 (2009).

\bibitem{moore07}
J.\ E.\ Moore and L.\ Balents,
Phys.\ Rev.\ B \textbf{75}, 121306(R) (2007).

\bibitem{bernevig06}
B.\ A.\ Bernevig, T.\ L.\ Hughes, and S.-C.~Zhang,
Science \textbf{314}, 1757 (2006).

\bibitem{konig07}
M.\ K\"onig \textit{et al.},
Science \textbf{318}, 766 (2007).

\bibitem{Fu06_3Da}
L.\ Fu, C.\ L.\ Kane, and E.\ J.\ Mele,
Phys.\ Rev.\ Lett.\ \textbf{98}, 106803 (2007).


\bibitem{Qi2008}
X.-L.\ Qi, T.\ Hughes,  and S.-C.\ Zhang,
Phys.\ Rev.\ B \textbf{78}, 195424 (2008).

\bibitem{hasan07}
D.\ Hsieh \textit{et al},
Nature \textbf{452}, 970 (2008).


\bibitem{SRFL}
A.\ Schnyder, S.~Ryu, A.~Furusaki, A.\ Ludwig,
% `` Classification of topological insulators and superconductors in three spatial dimensions,''
Phys.\ Rev.\ B \textbf{78}, 195125 (2008);
%[arXiv:0803.2786];{
%``Classification of Topological Insulators and Superconductors'',
AIP Conf. Proc. \textbf{1134}, 10 (2009);
%[arXiv:0905.2029].
S.\ Ryu, A.\ Schnyder, A.~Furusaki, and A.\ Ludwig,
\texttt{arXiv:0912.2157}.
%New J.\ Phys.\, to be published (2009).



\bibitem{Kitaev}
A.~Kitaev,
%``Periodic table for topological insulators and superconductors,''
AIP Conf.\ Proc. \textbf{1134}, 22 (2009).


\bibitem{Horava}
The importance of K-theory in the physics of Fermi surfaces has been
first pointed out in P.~Horava,
  %``Stability of Fermi surfaces and K-theory,''
  Phys.\ Rev.\ Lett.\  {\bf 95}, 016405 (2005).
% [arXiv:hep-th/0503006].
  %%CITATION = PRLTA,95,016405;%%



\bibitem{WittenK}
  E.~Witten,
%  ``D-branes and K-theory,''
  JHEP {\bf 9812}, 019 (1998);
%  [arXiv:hep-th/9810188].
  %%CITATION = JHEPA,9812,019;%%
 P.~Horava, Adv.\ Theor.\ Math.\ Phys.\  {\bf 2}, 1373 (1999).


\bibitem{Sen}
  A.~Sen,
  %``Tachyon dynamics in open string theory,''
  Int.\ J.\ Mod.\ Phys.\  A {\bf 20}, 5513 (2005).
%  [arXiv:hep-th/0410103].
  %%CITATION = IMPAE,A20,5513;%%



\bibitem{Karch}
For a different approach to TIs from string theory, see, for example,
  A.~Karch,
  %``Electric-Magnetic Duality and Topological Insulators,''
  Phys.\ Rev.\ Lett.\  {\bf 103}, 171601 (2009).
  %[arXiv:0907.1528 [cond-mat.mes-hall]].
  %%CITATION = PRLTA,103,171601;%%


\bibitem{RTTA}
 S.~Ryu and T.~Takayanagi,
 \texttt{arXiv:1007.4234}.
  %%CITATION = ARXIV:1007.4234;%%



\bibitem{Wen91-99}
X.\ -G.\ Wen, Mod.\ Phys.\ Lett.\ B \textbf{5}, 39 (1991);
Phys.\ Rev.\ B \textbf{60}, 8827 (1999).
%81. arXiv:cond-mat/9811111 [ps, pdf, other]
%Title: Projective Construction of Non-Abelian Quantum Hall Liquids


\bibitem{Kitaev05}
%14. arXiv:cond-mat/0506438 [ps, pdf, other]
%Title: Anyons in an exactly solved model and beyond
Alexei Kitaev,
Ann.\ of Phys.\ \textbf{321}, 2 (2006).

\bibitem{Ryu08}
%24. arXiv:0811.2036 [ps, pdf, other]
%Title: Three-dimensional topological phase on the diamond lattice
Shinsei Ryu,
Phys.\ Rev.\ B \textbf{79}, 075124 (2009).


\bibitem{Rey}
  S.~J.~Rey,
  %``String Theory on Thin Semiconductors: Holographic Realization of Fermi
  %Points and Surfaces,''
  Prog.\ Theor.\ Phys.\  {\bf 177}, 128 (2009).
%[arXiv:0911.5295 [hep-th]]
  %%CITATION = ARXIV:0911.5295;%%

\bibitem{DKS}
  J.~L.~Davis, P.~Kraus and A.~Shah,
%  ``Gravity Dual of a Quantum Hall Plateau Transition,''
  JHEP {\bf 0811}, 020 (2008).
%  [arXiv:0809.1876 [hep-th]]
  %%CITATION = JHEPA,0811,020;%%

\bibitem{FujitaFQHE}
%8. arXiv:0901.0924 [ps, pdf, other]
M.~Fujita, W.~Li, S.~Ryu, and T.~Takayanagi,
JHEP \textbf{0906} 066, (2009).



\bibitem{ISS}
For a KR theory analysis in similar brane systems, see
T.~Imoto, T.~Sakai and S.~Sugimoto,
%  ``O(N) and USp(N) QCD from String Theory,''
\texttt{arXiv:0907.2968},
%%CITATION = ARXIV:0907.2968;%%
and references therein.







\end{thebibliography}
\end{document}